\documentclass[journal=nalefd,manuscript=letter]{achemso}

\usepackage[T1]{fontenc}       
\usepackage[utf8]{inputenc}
\usepackage{gensymb}
\usepackage{textcomp}
\DeclareUnicodeCharacter{00A0}{~}

\author{Toma Susi}
\affiliation[University of Vienna]
{University of Vienna, Faculty of Physics, Boltzmanngasse 5, 1090 Vienna, Austria}
\email{toma.susi@univie.ac.at}
\author{Trevor P. Hardcastle}
\affiliation[University of Leeds (SCaPE)]{Institute for Materials Research, SCaPE, University of Leeds, LS2 9JT, UK}
\author{Hans Hofsäss}
\affiliation[University of Göttingen]{II Institute of Physics, Faculty of Physics, University of Göttingen, Friedrich-Hund-Platz 1, 37077 Göttingen, Germany}
\author{Andreas Mittelberger}
\author{Timothy J. Pennycook}
\author{Clemens Mangler}
\affiliation[University of Vienna]
{University of Vienna, Faculty of Physics, Boltzmanngasse 5, 1090 Vienna, Austria}
\author{Rik Drummond-Brydson}
\affiliation[University of Leeds (ChemPro)]{School of Chemical and Process Engineering, Faculty of Engineering, University of Leeds, LS2 9JT, UK}
\author{Andrew J. Scott}
\affiliation[University of Leeds (SCaPE)]{Institute for Materials Research, SCaPE, University of Leeds, LS2 9JT, UK}
\author{Jannik C. Meyer}
\author{Jani Kotakoski}
\affiliation[University of Vienna]
{University of Vienna, Faculty of Physics, Boltzmanngasse 5, 1090 Vienna, Austria}
\email{jani.kotakoski@univie.ac.at}
\title[]{Single-atom spectroscopy of phosphorus dopants implanted into graphene}

\abbreviations{STEM, EELS, DFT}

\begin{document}



\begin{abstract}
One of the keys behind the success of the modern semiconductor technology has been the ion implantation of silicon, which allows its electronic properties to be tailored. For similar purposes, heteroatoms have been introduced into carbon nanomaterials both during growth and using post-growth methods. However, due to the nature of the samples, it has been challenging to determine whether the heteroatoms have been incorporated into the lattice as intended, with direct observations so far being limited to N and B dopants, and incidental Si impurities. Furthermore, ion implantation of these materials is more challenging due to the requirement of very low ion energies and atomically clean surfaces. Here, we provide the first atomic-resolution imaging and electron energy loss spectroscopy (EELS) evidence of phosphorus atoms incorporated into the graphene lattice by low-energy ion irradiation. The measured P~\textit{L}-edge response of an single-atom EELS spectrum map shows excellent agreement with an \textit{ab initio} spectrum simulation, conclusively identifying the P in a buckled substitutional configuration. Our results demonstrate the viability of phosphorus as a lattice dopant in $sp^2$-bonded carbon structures and provide its unmistakeable fingerprint for further studies.
\end{abstract}

The implantation of crystalline silicon with ions of boron, phosphorus or arsenic forms the foundation of the modern semiconductor industry, and is largely responsible for the proliferation of computing in the modern world. However, the limits of miniturization with this material are being reached, prompting great interest in nanomaterial alternatives such as single-walled carbon nanotubes (SWCNTs) and graphene. Both have superb intrinsic properties, but also challenges: nanotubes are produced as a mixture of semiconducting and metallic species,\cite{Saito92APL} whereas graphene lacks a band gap.\cite{Novoselov05N}

Great effort has been directed to controlling the electronic properties of these novel materials. Over the last decade, significant progress has been made in the purification and separation of nanotube samples,~\cite{Arnold06NN} and notable successes reached in their incorporation into electronics.\cite{Shulaker13N} In the case of graphene, efforts have been directed into opening a gap and to tuning the carrier concentration, for example by cutting graphene into nanoribbons~\cite{Jia11NS}, via strain~\cite{Si16NS}, by building van der Waals stacks~\cite{Geim13N}, and via chemical functionalisation~\cite{Yong-Jin152D}. Doping the structure with heteroatoms, either by introducing a precursor during growth or by post-growth processing such as ion implantation, is a particularly prominent route of the latter kind for both nanotubes and graphene.\cite{Ayala10RMP,Susi2015Book}

A commonly used tool for studying heteroatom doping is X-ray photoelectron spectroscopy, since the core level binding energies it measures are fingerprints of different chemical species.\cite{Susi15BJN} Unfortunately, the very low amount of dopant atoms corresponding to even relatively high concentrations, along with the synthesis byproducts and contamination inevitably present, make it very difficult for macroscopic characterization techniques to conclusively prove the incorporation of dopants into the lattice. Only when using carefully purified materials can there be a high degree of confidence that the spectroscopic signatures originate from heteroatoms in the lattice itself,\cite{Ruiz-Soria15C} but even then it is challenging to tease out their exact bonding, which is only possible by comparison to known references or simulations.\cite{Susi14BJN}

Scanning tunneling microscopy is a powerful tool for local characterization, and even though it lacks direct chemical sensitivity, it has been used to confirm the local bonding of N and B heteroatoms in graphene,\cite{Zhao13NL} and N in SWCNTs.\cite{Tison13AN} Recent advances in aberration-corrected scanning transmission electron microscopy\cite{Krivanek10N} (STEM) have similarly enabled the identification of individual atoms in low-dimensional materials such as graphene.\cite{Geim07NM} When atomic resolution STEM is used for electron energy loss spectroscopy\cite{Suenaga10N,Krivanek14N} (EELS), even the precise nature of the atoms' bonding can now be resolved with the help of first principles simulations.\cite{Suenaga10N,Nicholls13AN}. Hitherto, this method has been used to confirm lattice doping with both nitrogen and boron\cite{Arenal14NL, Bangert13NL, Kepaptsoglou15AN} as well as the lattice incorporation of the ubiquitous contaminant, silicon\cite{Zhou12PRL,Ramasse13NL,Susi14PRL}. However, such direct evidence for the lattice doping of $sp^2$-bonded carbon with any other element has so far been lacking.

Although phosphorus (P) was already early on proposed as a possible electronic donor,\cite{Strelko00C} the first experimental reports on doping graphitic materials with it were published relatively recently.\cite{Krstic10AN,Ruiz-Soria15C} Like nitrogen, phosphorus has five valence electrons, but on the third electron shell, yielding has a significantly larger covalent atomic radius (106~pm, compared to 82~pm for B, 77~pm for C, and 75~pm for N). Based on density functional theory (DFT) simulations, it is expected that P will predominantly bond to three C neighbors, but buckle significantly out of the plane,\cite{Cruz-Silva09AN} similar to Si where the specroscopic signature of this buckling was unambiguously identified.\cite{Ramasse13NL}

Krsti\v{c} and co-workers further suggested that P substitutions are readily oxidized in ambient, with the P--O bond formation predicted to be exothermic by as much as 3.3~eV.\cite{Krstic10AN} This found recent support from a study of carefully purified P-doped single-walled carbon nanotubes, which found a decrease in the x-ray photoelectron spectroscopy signature corresponding to oxidized P upon annealing.\cite{Ruiz-Soria15C} However, even though these samples were a significant advance over previous studies, direct evidence for the incorporation of phosphorus into the lattice of any carbon nanomaterial has so far been missing.

To address this, we implanted low-energy P ions into commercial monolayer graphene (Quantifoil\textregistered~R~2/4, Graphenea) at a mass-selected ion beam deposition system.\cite{Bangert13NL} Before inserting the samples into the deposition chamber, they were baked on a hot plate in air at 400$\degree$C for 15~min in an attempt to reduce contamination. The source of phosphorus was a hot filament hollow cathode plasma ion source (Model SO-55, High Voltage Engineering) with an oven containing a small amount of red phosphorus. For the implantation, a 30~keV mass-selected $^{31}$P$^+$ ion beam was first deflected to eliminate any neutralized ions and decelerated toward the sample. The deceleration bias voltage was set relative to the ion source anode potential, resulting in a maximum ion energy of 30~eV (with a few eV tail toward lower energies). The samples were irradiated in a 2$\times$10$^{-6}$~Pa vacuum at room temperature with a fluence of (4$\pm$1)$\times$10$^{14}$~cm$^{-2}$.

The ion energy of 30 eV was chosen in an effort to obtain substitutions without causing significant damage. Based on the conservation of momentum and energy, $^{31}$P with a kinetic energy of 30~eV can transfer a maximum of 26.22~eV to $^{12}$C in a head-on collision. The displacement threshold energy is the minimum energy required to remove and atom from the material, and in graphene it is graphene is 21.14~eV (ref. \citenum{Susi16NC}). The energy remaining after an impact should not be enough for the P ion itself to escape, especially considering that for most impact parameters the transferred energy is lower than the maximum. Thus this ion energy should be a conservative estimate.

\begin{figure}
\includegraphics[width=0.8\textwidth]{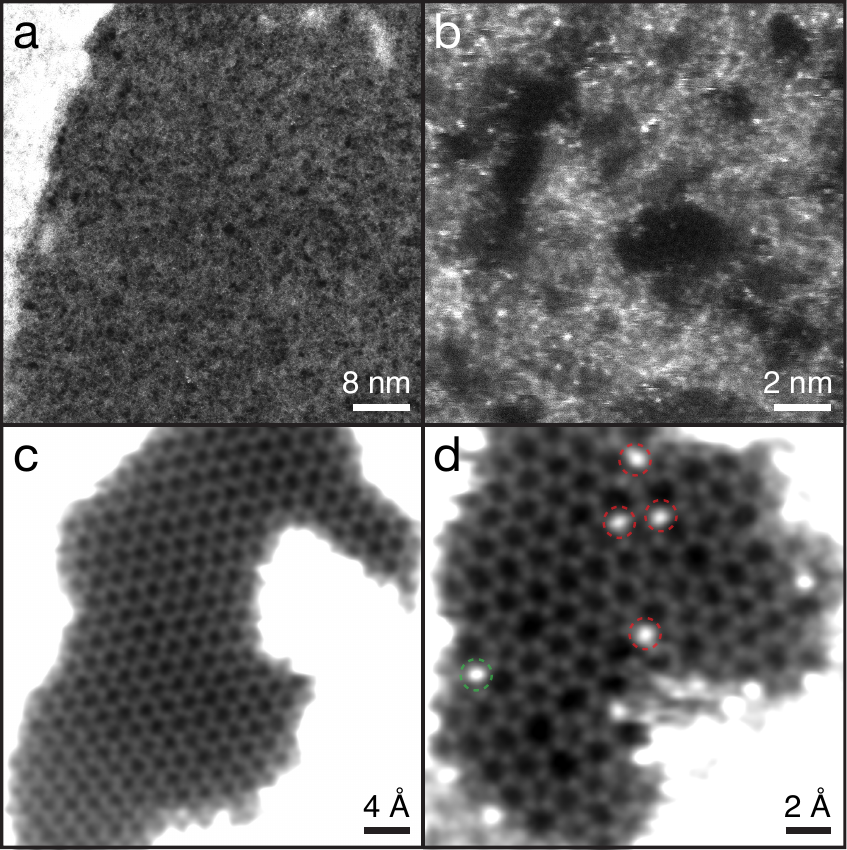}
\caption{Overview STEM/MAADF images (1024$\times$1024~px) of the graphene sample implanted with 30~eV $^{31}$P$^+$ ions. a) Throughout the sample, the graphene surfaces were almost completely covered by contamination. b) Despite extensive searching, the largest clean areas that we found were only a few nm in size. The contamination contains many heavier atoms. c) Of the clean monolayer areas, most did not contain any discernible dopants or defects. d) In some cases, disordered areas with many heavier atoms incorporated into the lattice could be found. However, most are Si (red dashed circles) instead of P (green dashed circle) (see text). Panels c and d have been treated for clarity by Gaussian blurs with radii of 8 and 5~px, respectively.\label{STEM}}
\end{figure}

We observed the samples in a Nion UltraSTEM100 microscope operated at 60~keV in near-ultrahigh vacuum with a beam convergence semiangle of 30~mrad. As a general observation, the sample surfaces were very contaminated (Figure~\ref{STEM}a), with the largest clean areas that we could find only a few nanometers in size (Figure~\ref{STEM}b). This could be either due to contamination in the ion deposition chamber, or due to the pinning of atmospheric contamination on chemically reactive dopant and defect sites. Of the areas that could be imaged, most contained no discernible dopants (Figure~\ref{STEM}c). However, we did find several small clean areas with heavier atoms incorporated into the lattice, including a slightly disordered area with several dopants (Figure~\ref{STEM}d) and one clear instance of a single substitution we will discuss later.

In STEM, the annular dark field contrast is directly sensitive to atomic number in so-called Z-contrast.\cite{Krivanek10N} However, in the case of phosphorus dopants, the situation is complicated due to the ubiquitous presense of silicon contaminants,\cite{Ramasse13NL,Susi14PRL} which have almost the same number of protons. While the contrast difference is detectable, it can be challenging to discriminate between the two atomic species without spectroscopy. Our EELS acquisition setup consists of a Gatan PEELS~666 spectrometer retrofitted with an Andor iXon~897 electron-multiplying charge-coupled device (EMCCD) camera. A spectrum image can be acquired by sending a synchronization signal from the Nion Swift software to the camera via a custom-installed Rasperry Pi minicomputer. The energy dispersion was 0.722~eV/pixel in the raw data (two-point calibration using the Si~\textit{L}$_{23}$-and C~\textit{K}-edges measured separately over a three-coordinated Si atom.\cite{Ramasse13NL}) Our energy resolution is around 0.4~eV, the beam current was close to 30~pA, and the EELS collection semiangle was 35~mrad.\\

We were able to identify the five heteroatoms of Figure \ref{STEM}d simply by comparing their relative intensities to a quantitative STEM image simulation graphene with separated single Si and P atoms. This was created with the QSTEM software package~\cite{Koch2002} using our instrumental parameters (chromatic aberration coefficient 1~mm, energy spread 0.3~eV; spherical aberration coefficient 1~$\mu$m; thermal diffuse scattering included via frozen phonon modeling at 300~K; additional instabilities (such as sample vibration) simulated by blurring the resulting image (Gaussian kernel with a sigma of 0.39~\AA); and the medium-angle annular dark-field detector angle range set to the experimental range of 60--80 mrad).

\begin{figure}[t]
\includegraphics[width=0.65\textwidth]{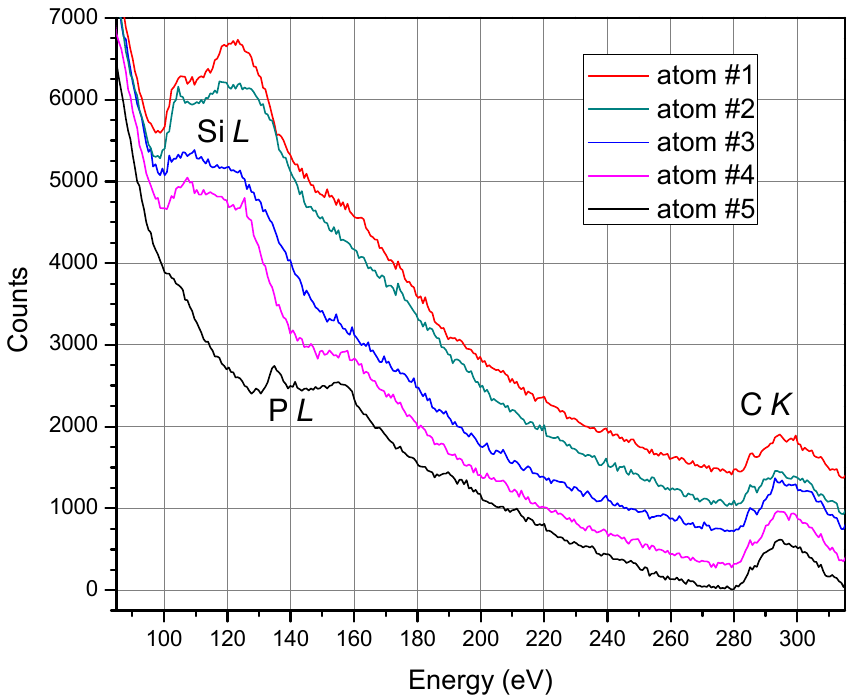}
\caption{Spot EELS spectra (0.5~s acquisition time) measured over five bright atoms incorporated into the graphene lattice (Figure~\ref{Psub}d). The first four spectra show a clear Si~\textit{L}-edge, while only the fifth one displays a P~\textit{L}-edge instead. The spectra have been slightly offset for clarity.\label{SpotEELS}}
\end{figure}

From the simulation, we find that P is expected to be 1.11~times brighter than Si, corresponding to a Z-contrast of approximately Z$^{1.71}$. The four atoms marked by red dashed circles in Figure~\ref{STEM} have relative intensities of 1.000$\pm$0.023, whereas the one marked with the green dashed circle is brighter than the others with a relative intensity of 1.083$\pm$0.018, consistent with being P. Notably, this intensity cannot correspond to oxidized P, but since a 60~keV electron can transfer up to 8.7~eV to an O atom in a head-on collision, we would not expect it to stay bound under the intense electron irradiation (similar to O in graphene oxide\cite{Tararan16CM}). EELS spectra measured over these atoms are shown as Figure~\ref{SpotEELS}, also clearly indicating that four of the atoms are Si and only one is P. Unfortunately the atomic configuration of this disordered patch was not stable due to beam-induced bond rotations,\cite{Kotakoski11PRB,Susi14PRL} preventing us from capturing higher quality spectra.

The instability of Si and P heteroatoms is not surprising, since 60 keV electrons can transfer enough energy to the C atoms neighboring them to cause displacements or bond rotations. For Si, we previously calculated the C neighbor displacement threshold energy to be in the range $[16.75,17.00]$ eV.\cite{Susi14PRL} Using the same density functional theory (DFT) molecular dynamics methodology, we find the same threshold for displacing a C next to a substitutional P. These thresholds are for momentum transfers perpedicular to an otherwise perfect graphene lattice, and are thus very likely overestimates for a disordered area. Both Si and P are too heavy to be directly displaced by the beam.

\begin{figure}[t]
\includegraphics[width=0.65\textwidth]{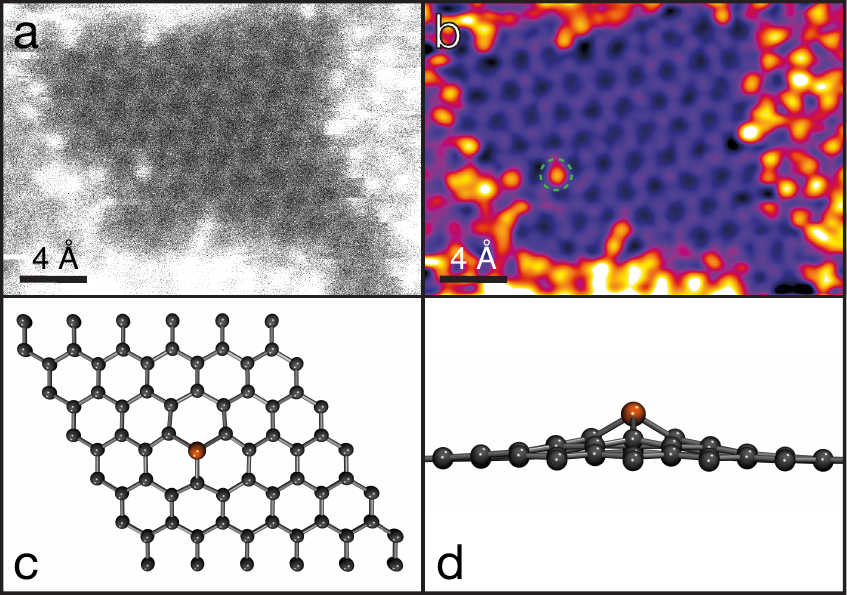}
\caption{A substitutional P atom in graphene. a) Cropped unprocessed STEM/MAADF image. b) The same image colored and filtered with a double Gaussian procedure\cite{Krivanek10N} ($\sigma_1=28.4$~px, $\sigma_2=6.4$~px, weight $= 0.22$), with the P atom indicated by the dashed green circle. c) Relaxed model structure of a P substitution in a 6$\times$6 graphene supercell. d) Side view of the model structure.\label{Psub}}
\end{figure}

\begin{figure}[t]
\includegraphics[width=0.65\textwidth]{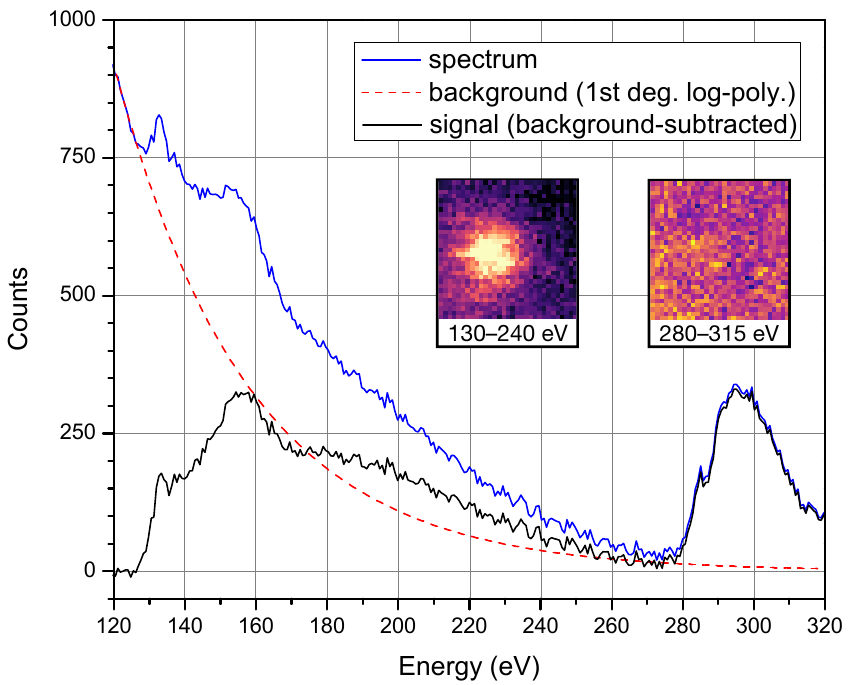}
\caption{EELS spectrum averaged from 17$\times$17 pixels over the substitutional P dopant (Figure \ref{Psub}a--b) shown in the overlaid spectrum map (energy windows from 130--240 eV and 280--315 eV for the P~\textit{L} and C~\textit{K}-edges, respectively, colored with the ImageJ lookup table `mpl-magma'). The blue line is the original spectrum, red is a background fit using a 1st degree log-polynomial, and black the resulting signal. The energy dispersion was 0.722~eV/px, and the spectrum binned over a $\sim$80$\times$512~pixel slice of the 512$\times$512~px EMCCD camera.\label{fullEELS}}
\end{figure}

We thus searched for a clearer example of a P substitution, and found a single bright atom in an otherwise pristine lattice (Figure \ref{Psub}a--b). DFT simulations indicate that apart from its brightness, a P substitution would appear very similar to C in the projected STEM image (Figure \ref{Psub}c). However, the P atom buckles 1.467~\AA~out of the graphene plane, resulting in P--C bond lengths of 1.759~\AA~(Figure \ref{Psub}d). To confirm the identity of this atom and its bonding, we recorded an EELS spectrum map with a dwell time of 50~ms per pixel for a total acquisition time of 51.2~s for the 32$\times$32~px map (Figure \ref{fullEELS}). Notably, the stage drift during the aquisition was practically non-existent. To subtract the low-loss background, we fitted a first degree log-polynomial\cite{Wilson91MMM} to the spectrum preceding the P edge. The P \textit{L}-edge starting at $\sim$130 eV is well localized on the atomic site, whereas the C \textit{K}-edge appears over the entire mapped area.

To simulate the EELS response of the P substitution,\cite{Gao08PRB-2} we used DFT implemented with self-consistently-generated on-the-fly pseudopotentials in the CASTEP package.\cite{Vackar98PRB} The structure was relaxed using a TS-corrected\cite{Tkatchenko09PR} PBE functional with a plane-wave cutoff energy of 600 eV and \textbf{k}-point spacings $<0.02$ \AA$^{-1}$ in a 6$\times$6 graphene supercell with a lattice parameter of 2.46668~\AA~and 20~\AA~of perpedicular vacuum.

The charge state of a phosphorus dopant has raised some debate. It has been suggested that P will either act as a net donor,\cite{Krstic10AN} or that it will bond in sp\textsuperscript{3} hybridization creating a nondispersive localized state.\cite{Cruz-Silva09AN} Based on Bader analysis\cite{Tang09JoPCM} of the all-electron density derived from our DFT simulation, the P is found to donate 1.79 electrons, with its three C neighbors receiving 1.68. Thus even at zero Kelvin there is indication of charge transfer to the lattice.

The P~\textit{L}- and C~\textit{K}-edge spectra were calculated by evaluating the perturbation matrix elements of the transitions from the P~2\textit{p} and C~1\textit{s} core states to the unoccupied states calculated up to 6000 bands. Notably, we used no explicit core hole,\cite{Susi15PRB} as this has been found to result in significantly better agreement with experimental spectra.\cite{Ramasse13NL,Kepaptsoglou15AN} The resulting densities of state were broadened using the OptaDOS package\cite{Nicholls12JoP} with a 0.4~eV Gaussian instrumental broadening and additional semi-empirical 1.26~eV Lorentzian lifetime broadening for the P~\textit{L}-edge and 0.17~eV for the C \textit{K}-edge. The theoretical spectra were then rigidly shifted along the energy axis to achieve the best fit and normalised to the experimental signal.

\begin{figure}[t]
\includegraphics[width=0.6\textwidth]{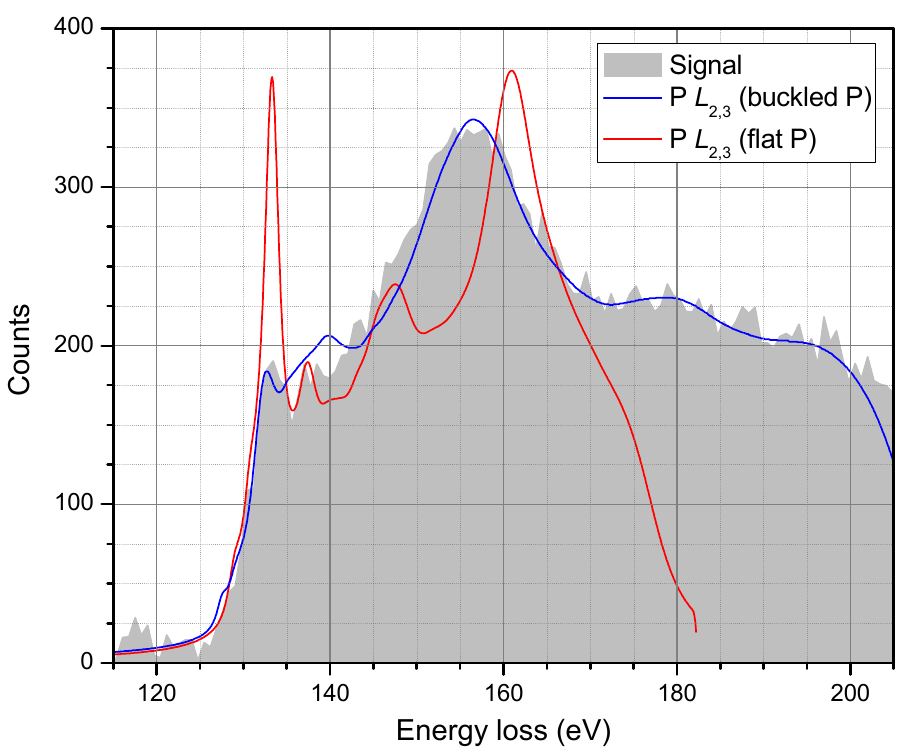}
\caption{The measured background-subtracted EELS signal (filled grey area) together with the simulated EELS response (colored lines). The P~\textit{L}$_{2,3}$ edge, with simulated spectra shown for the DFT-relaxed buckled P configuration (blue) and a metastable flat configuration (red).\label{EELSedges}}
\end{figure}

The close agreement between the simulated P~\textit{L}$_{2,3}$ spectrum and the experimental signal (Figure~\ref{EELSedges}a) proves that the measured atom is P in the buckled\cite{Ramasse13NL} substitutional configuration. The only disagreements are the slightly more sudden simulated onset and the absence of the small peak around 140~eV. (To conclusively rule out the metastable completely planar bonding, we did simulate its spectrum but found the $\pi^*$ peak to be dramatically overestimated and the maximum of the $\sigma^*$ response to be $\sim$5 eV too high in energy.) We also calculated the P~\textit{L}$_{1}$ response, but as the fit of Figure~\ref{EELSedges} is already excellent, its inclusion would dramatically overestimate the intensity starting from the edge onset around 180~eV. The dipole sum rule ($\Delta l = \pm 1$) is thus strictly enforced by our scattering geometry, as expected.\cite{Egerton14MRT}


Why were most of the heteroatoms we could observe in the lattice Si and not P? The ion beam itself is stricly mass-selected, and there is no Si in the ion source materials. One possible explanation is that the implanted P atoms are more chemically reactive than Si due to their extra valence electron, and thus more efficient in attracting the obscuring contamination. Another possibility would be that the P ions were mostly causing displacements instead of substitutions, which then get filled by the ubiquitous mobile Si contaminants present in graphene samples.\cite{Susi14PRL} Although unfortunately no interatomic P--C potential is available to estimate the optimal energy for achieving the highest probability of substitutions, for Si this was recently calculated using molecular dynamics simulations\cite{Li15RSCA} to be around 50--70 eV. Differences in the valence electron structure ($\Delta \mathrm{Z}=1$) and atomic mass ($\Delta \mathrm{A}=4$) notwithstanding, it would be quite surprising if P ions of only 30~eV would predominantly cause damage.

To conclude, we have implanted graphene with phosphorus ions, and shown that they bond in the expected buckled substitutional configuration. However, working with phosphorus is challenging, since contamination layers cover most of the lattice, containing Si atoms that can be easily mistaken for P. Further work is needed to optimize the implantation conditions and to clean the samples both post- and pre-deposition. Nonetheless, ion implantation is a feasible route to this novel doped material, and our simulated spectrum will serve as an unmistakeable fingerprint of the heteroatom for further chemical synthesis studies.

\begin{acknowledgement}
T.S. acknowledges funding from the Austrian Science Fund (FWF) via project P 28322-N36 and the Vienna Scientific Cluster for computational resources. A.M. and J.C.M. acknowledges funding by the FWF project I1283-N20, and J.C.M. and  C.M. by the European Research Council Grant No. 336453-PICOMAT. T.J.P. was supported by the European Union's Horizon 2020 research and innovation programme under the Marie Skłodowska-Curie grant agreement No. 655760 -- DIGIPHASE, and J.K. by the Wiener Wissenschafts-, Forschungs- und Technologiefonds (WWTF) via project MA14-009.
\end{acknowledgement}
%
%
%

\providecommand{\latin}[1]{#1}
\providecommand*\mcitethebibliography{\thebibliography}
\csname @ifundefined\endcsname{endmcitethebibliography}
  {\let\endmcitethebibliography\endthebibliography}{}

\end{document}